\documentclass[fleqn,10pt]{wlscirep}
\usepackage[utf8]{inputenc}
\usepackage[T1]{fontenc}
\usepackage{lineno}
\usepackage{adjustbox}
\usepackage{multirow}

\title{Empowering Open Data Sharing for Social Good: A Privacy-Aware Approach}

\author[1,*]{Tânia Carvalho}
\author[2]{Luís Antunes}
\author[3]{Cristina Costa}
\author[4]{Nuno Moniz}
\affil[1]{Faculty of Computer Science, University of Porto, Portugal}
\affil[2]{TekPrivacy, Porto, Portugal}
\affil[3]{Faculty of Medicine, University of Porto, Portugal}
\affil[4]{Lucy Family Institute for Data and Society, University of Notre Dame, Indiana, USA}

\affil[*]{corresponding author: tania.carvalho@fc.up.pt}

\begin{abstract}
The Covid-19 pandemic has affected the world at multiple levels. Data sharing was pivotal for advancing research to understand the underlying causes and implement effective containment strategies. In response, many countries have promoted the availability of daily cases to support research initiatives, fostering collaboration between organisations and making such data available to the public through open data platforms.
Despite the several advantages of data sharing, one of the major concerns before releasing health data is its impact on individuals' privacy. Such a sharing process should be based on state-of-the-art methods in {\it Data Protection by Design and by Default}.
In this paper, we use a data set related to Covid-19 cases in the second largest hospital in Portugal to show how it is feasible to ensure data privacy while improving the quality and maintaining the utility of the data. Our goal is to demonstrate how knowledge exchange in multidisciplinary teams of healthcare practitioners, data privacy, and data science experts is crucial to co-developing strategies that ensure high utility of de-identified data.
\end{abstract}
\begin{document}

\flushbottom
\maketitle

\thispagestyle{empty}

\section*{Introduction}

The free use, reuse and redistribution~\cite{opendata} makes open data invaluable in driving progress, fostering innovation and helping to solve humanitarian and societal challenges, as demonstrated in the context of Covid-19~\cite{cosgriff2020data}. Open data sharing enables broader participation in research studies and faster conclusions, as it is accessible to everyone. In addition, government legislation requiring transparency has led to an expansion of publicly available data resources. Several open government data initiatives have emerged in diverse countries, particularly for data related to the public sector. In particular, the European Union~\cite{eudirective}, and the U.S. Chief of Information Officer's Council~\cite{act2018} among others encourage governments to provide access to public sector information, promoting its reuse and accelerating innovation. 

The vast advantages of open access to data have led to transformative shifts. However, with this came the emergence of privacy threats, prompting legislators to develop regulatory measures to ensure the responsible and ethical use of data. 
Therefore, regulations such as General Data Protection Regulation\footnote{https://eur-lex.europa.eu/eli/reg/2016/679/oj} (GDPR) and California Consumer Privacy Act\footnote{https://oag.ca.gov/privacy/ccpa} (CCPA), emerged to balance the facilitation of open data with the protection of individual privacy rights, ultimately contributing to innovation and progress while safeguarding against potential risks.
In response to these developments, many organisations, institutions and governments endorse the FAIR (Findable, Accessible, Interoperable and Reusable) guiding principles~\cite{wilkinson2016fair, fair_sharing}. FAIR describes considerations for exploration, sharing and reuse in data publishing environments. 

Besides privacy concerns, multiple challenges have been identified, such as benefits and costs, legislative issues, data formats and infrastructure limitations, among others~\cite{kucera2015methodologies}. Furthermore, a large amount of collected data is often unused as it is redundant or obsolete. The gathering of data should follow the principles of data protection by design and by default approach~\cite{privacy_by_design}. 
Article 25 of the GDPR requires data controllers to implement appropriate technical and organisational measures to ensure data protection principles are integrated into processing activities from the outset (data protection by design) and that, by default, only the necessary personal data for each specific purpose is processed (data protection by default). This approach enables organisations to optimise data storage efficiency, improving sustainability and bolstering their environmental footprint~\cite{alessi2021privacy}.

In light of the numerous benefits and associated privacy risks of public dissemination of data, it is imperative to protect the privacy of individuals while maximising the data utility for diverse applications. Achieving an optimal trade-off between these conflicting goals is a challenge, amplified by the scarcity of professionals with the required expertise in both fields. Therefore, to achieve optimal outcomes, it is essential to assemble a multidisciplinary team dedicated to balancing privacy preservation and utility maximisation. 

During the pandemic, the second largest hospital in Portugal received a large number of data access requests for clinical research purposes. Given the high interest in accessing Covid-19 data, the hospital created a team of experts with a heterogeneous set of skills to assess the possibility of providing a de-identified data set with high quality and maintaining its utility for the majority of the requests. Therefore, this work was conducted in collaboration with the hospital. We provided recommendations for the transformation of the Covid-19 patient data set from the hospital in accordance with data protection principles and standards. While related works focus on transforming data following de-identification algorithms~\cite{bandara2020evaluation, jakob2020design} such as $k$-anonymity~\cite{samarati2001protecting}, our approach involves safeguarding the data set through a customised transformation guided by the experts' input.

The main purpose of this paper is to demonstrate the importance of healthcare practitioners' insights to enhance domain understanding, thereby enabling the application of appropriate privacy-preserving techniques (PPTs), and at the same time the importance of increasing healthcare professionals' awareness of privacy-enhancing technologies.  
We comprehensively detail the steps of the de-identification process, aiming to achieve an optimal balance between mitigating privacy risks and preserving data utility. Most importantly, we provide a high-quality and statistically useful data set that is privacy-compliant for sharing for social good. We are committed to fostering understanding, innovation and the development of tools that are critical to addressing the challenges posed by the pandemic.


\section*{Privacy-Aware Open Data}
In the context of privacy-aware, appropriate governance mechanisms should be established to comply with privacy and cybersecurity laws. 
Under Article 35 of the GDPR, a Data Protection Impact Assessment (DPIA) is mandatory when a new processing activity is likely to result in a high risk to the rights and freedoms of natural persons. A DPIA is a procedure that should be employed as an early cautioning framework as it identifies potential privacy risks. Several problems can be detected during the design of the project, such as the inappropriate legal grounds for processing and the exercise of data subject rights. When there is a need to design an (pseudo)anonymisation protocol, a suitable framework is the \textit{Five Safes}~\cite{arbuckle2019five, arbuckle2020building}. Such a framework aims to define (pseudo)anonymisation protocols and how rigorously the protocol should be. The \textit{Five Safes} framework is usually used as a privacy risk assessment and support to the decision-making. Decision trees~\cite{zipper2019balancing} are also used to evaluate practices for data sharing which help researchers in the process.

The urgency of data sharing has become even more crucial, particularly in light of the evolving Covid-19 pandemic, where timely access to multiple data sources was paramount due to the rapid dynamics of the pandemic and the fact that many case reports quickly became outdated.
Such resources, coupled with data-driven methodologies, facilitate enhancements in the response to control the spread of the virus. For instance, it was demonstrated that the shortage of face masks can be effectively managed through open government data in conjunction with online services~\cite{kim2020lesson}.
Several initiatives have been created to provide up-to-date Covid-19 information, e.g. Google Health Covid-19 Open Data Repository~\footnote{https://health.google.com/covid-19/open-data/}, Facebook Data For Good Covid-19~\footnote{https://dataforgood.facebook.com/dfg/covid-19} and Data Science for Social Good Portugal~\footnote{https://github.com/dssg-pt/covid19pt-data}. However, some limitations and difficulties were found in most of the available Covid-19 data~\cite{alamo2020covid}. These challenges include inconsistencies in data formats, outdated reports, difficulties in estimating mortality, and lack of information at the individual record level. 

Within the scope of Covid-19, public health authorities and employers are allowed by national law to process personal data under the conditions set by the EDPB (European Data Protection Board)~\cite{edpb_covid}.

Since the beginning of the pandemic, several research groups in Portugal have expressed a strong desire to contribute to the production of useful evidence for decision-making. To support these efforts, they requested access to healthcare data related to Covid-19. In response, the DGS (Portuguese Directorate-General of Health) released to the public a document describing the data set to be provided and a form to be filled in by researchers interested in requesting this data set. By submitting a study protocol approved by an ethics committee, researchers received this data set. However, the data set proved to be of limited value due to poor data quality~\cite{costa2021covid} and raised concerns about patient re-identification with minimal effort~\cite{carvalho2021fundamental}.

\subsection*{Risk-Benefit Evaluation}

Identifiability plays a key role in privacy laws and regulations. Basic frameworks provide an understanding of how an IT system processes personal data, which helps such systems comply with the Privacy by Design definition. For many organisations, de-identification starts from the stored identified data. However, sharing data within the same organisation but across different departments or with external entities poses multiple challenges. 
In this section, we introduce \textit{Five Safes} framework~\cite{arbuckle2019five} that helps to evaluate the risks and benefits of releasing or sharing a data set.

\subsubsection*{Risk Evaluation}
\begin{figure}[!ht]
   \centering
   \scriptsize
   \includegraphics[width=0.35\linewidth]{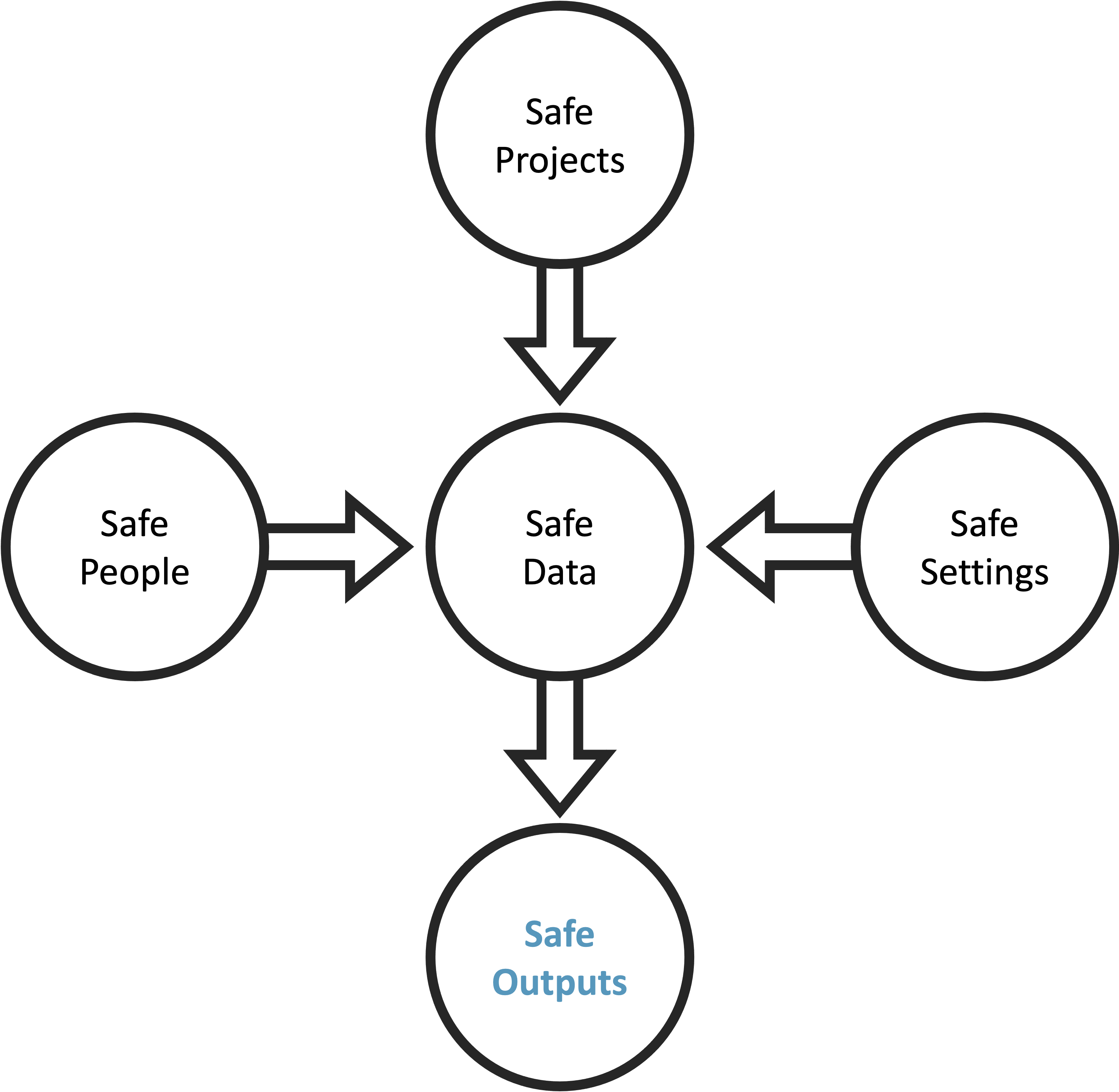}
 \caption{\textit{Five Safes} framework based on privacy risks.}
 \label{fig:5safes}
\end{figure}

\textit{Five Safes} framework has been used by organisations for more than a decade as it ensures that data utility is proportional to data privacy. As this framework is a privacy risk assessment and an auxiliary to the decision-making, it is recommended for researchers whether in the healthcare field, economics or commerce. The release could be directly made for researchers under certain circumstances and criteria or publicly available for any subject as open data. Figure~\ref{fig:5safes} summarises each step of \textit{Five Safes} framework. 

In each stage of the framework, there is a set of data transformations and technical and administrative controls that help to mitigate the risks. We briefly review the goals that we should attend for each stage. 

\begin{itemize}
    \item \textit{Safe Projects}: Understand the data flow to identify the legal and ethical limits of sharing personal information. Considerations include the origin of the data, recipient access requirements, and methods for identifying and protecting personal data.
    \item \textit{Safe People}: Besides the primary recipient, others may access the data, raising concerns about unintentional or intentional re-identification. To prevent a re-identification, we must assume that potential attackers have access to data and some background knowledge. 
    \item \textit{Safe Settings}: The data environment must be secure to prevent skilled attackers from re-identifying individuals. Standard controls include access restrictions, collaboration agreements, retention policies, and personal data handling should be accountable and transparent. 

    \item \textit{Safe Data}: After considering the people and environment in which the data is managed, the probability of attack and re-identification in the event of a successful attack must be assessed. The evaluation of the context and the data allows the identification of the threads to be managed.

    \item \textit{Safe Outputs}: Although the original purpose of de-identified data was established, recipients may draw new conclusions. Careful consideration is needed to avoid misusing such insights. Defining a privacy threshold based on context and data evaluation is essential.
 
\end{itemize}

\subsubsection*{Benefit Evaluation}
In addition to assessing the privacy risks of a data-driven project, it is fundamental to understand whether the data that will be shared or released to the public will bring benefits to those who access it. In general, benefits are evaluated according to the usability/utility of specific or open datasets.

The benefits of sharing data between organisations, or between departments within the same organisation, include reducing redundant effort, improving collaboration, and enabling effective use of resources. In the case of open data, the benefits can be social, environmental, health, political, economic and many others. These benefits can improve government transparency and public engagement, stimulate innovation, optimise government processes, and enable public/private data merging.   

When assessing the benefits, it is important to be aware of the privacy risks when collecting, processing and sharing data. Therefore, to incorporate the benefits into the risk-benefit assessment, the following considerations should be made~\cite{green2017open}: the attributes that can provide positive outcomes for an organisation, the forms in which the attributes will be beneficial, the possible recipients who will use the data, the likelihood of successful benefits in using the data, and the positive effects on people.

\subsubsection*{Risk-Benefit Matrix}
It is essential to assess the risks and benefits of data sharing, as it may result in the release of useless data or it may result in the release of data posing some risks to the data subject. The legitimate benefits to individuals or society and the potential harms of inappropriate data processing need to be weighed up beforehand. It is also essential to ensure the adequacy of the data subject's consent to disclosure.

Green et al.~\cite{green2017open} present an approach to privacy preservation that can be adopted by cities and other stakeholders releasing data to the public. The authors focus on four main recommendations: \textit{i)} risk-benefit analyses to further inform the development of open data projects, \textit{ii)} implementing appropriate safeguards in each stage of the data lifecycle, \textit{iii)} developing operational structures and processes that focus on privacy mitigation, and \textit{iv)} engaging the public as an essential aspect of data management.

\begin{figure}[ht!]
  \centering
    \includegraphics[width=0.35\textwidth]{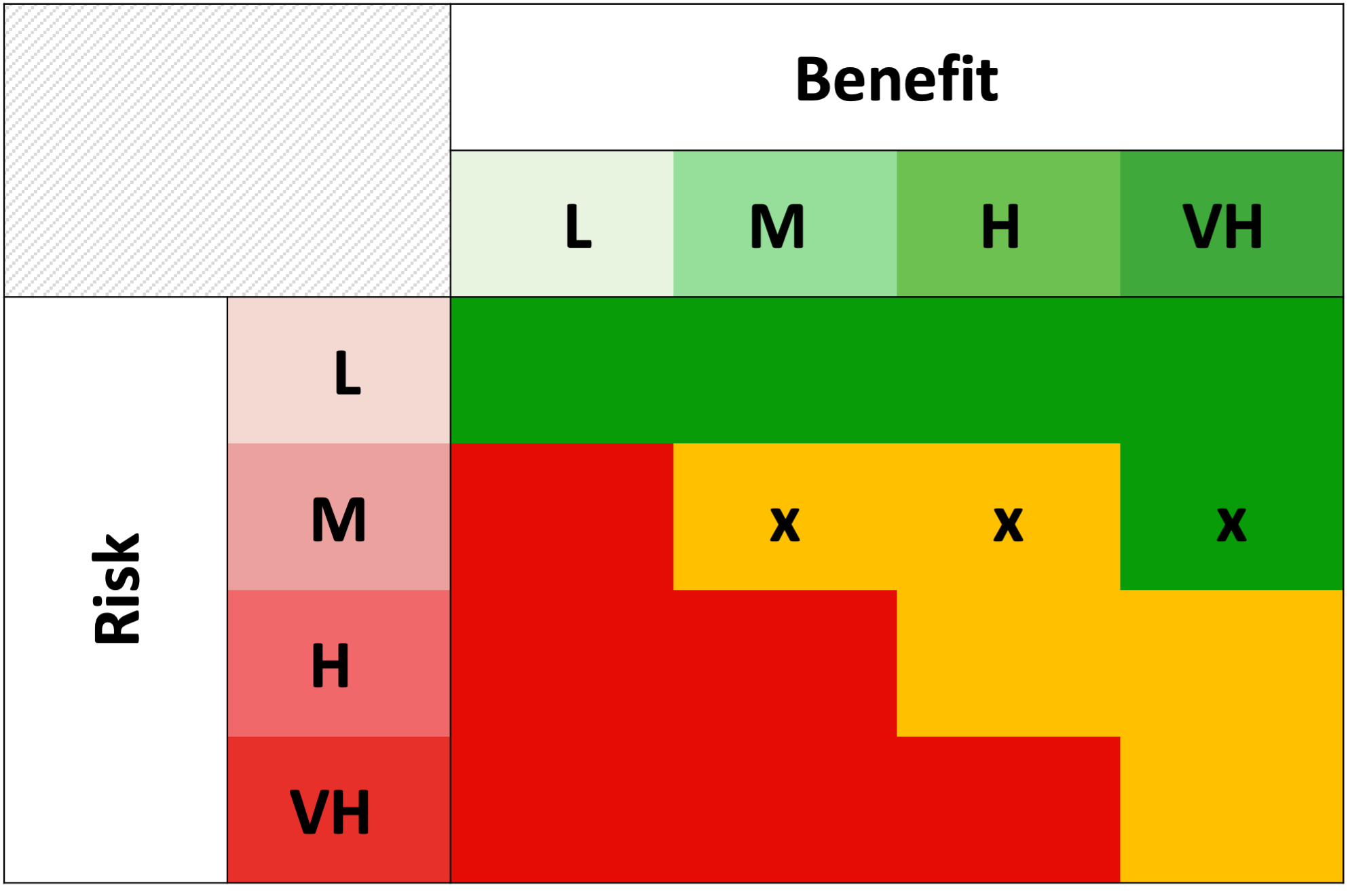}
    \caption{Example of a risk-benefit matrix.}
    \label{fig:risk_benefit}
\end{figure}

There are different matrices for representing risk-benefit levels, which help to understand at first-hand whether a data set could be publicly available. Figure~\ref{fig:risk_benefit} shows an example with four levels, namely low (L), moderate (M), high (H) and very high (VH). 
The most common level of publication is moderate (represented by x ticks). If the benefit is very high, it should be considered whether the risks outweigh the release of the data being analysed. It should be noted that the benefit is calculated at the asset level, not at the data set level. 

\section*{Methods}
To ensure \textit{safe data}, we apply a privacy mechanism to protect individuals' personal information and provide sufficient granularity for useful and meaningful further data analysis. One such mechanism is the \textbf{de-identification process}~\cite{carvalho2023survey}. Note that, the decision on the appropriate level of de-identification usually depends on the purpose of the data processing. 

The de-identification process consists of four main phases: \textit{i)} attributes classification, \textit{ii)} raw disclosure risk and data utility assessment, \textit{iii)} application of PPTs based on disclosure risk, data structure and attributes' characteristics, and \textit{iv)} re-assessment of disclosure risk and data utility. If there is no balance between the two measures, the transformation techniques should be refined; otherwise, the de-identified data can be shared. 

Quantifying the risk of disclosure is challenging since disclosure of confidential information generally occurs when an attacker has external information that the data controller often cannot anticipate.
Typically, an attacker is someone who has the skills, resources and motivation to re-identify data subjects or derive new information about them from de-identified data. If an attacker has more background knowledge than assumed, the risk of disclosure may be underestimated. Therefore, the controller needs to make prudent assumptions about such knowledge to predict the risk of disclosure. Typically, the controller determines the privacy risk under different scenarios (threat models), e.g. different sets of attributes that attackers may know.

Such attributes are classified as follows: \textbf{direct identifiers} which directly identify an individual such as name and social security number attributes, \textbf{quasi-identifiers (QI)} which may lead to identification when combined such as date of birth, gender and profession, and \textbf{sensitive attributes} which correspond to highly critical attributes usually protected by law and regulation such as religion, health information and ethnic group. 

Once the attributes classification has been established, the risk of disclosure, often referred to as the risk of re-identification, is assessed. We use the two most popular measures, $k$-anonymity and record linkage, to assess singling out and linkability respectively. $K$-anonymity~\cite{samarati2001protecting}, indicates how many $k$ occurrences appear in the data set w.r.t a set of QI values. 
If $k=1$, an attacker can single out the corresponding individual as the record is unique in the data set. For better protection, $k$ must be greater than 1. Record linkage~\cite{fellegi69}, measures the re-identification ability by linking two records. 
Given two data sets, \textit{A} (original) with \textit{a} elements and \textit{B} (de-identified) with \textit{b} elements, the resulting comparison space is the product of all possible pairs of records ($A \times B = \{(a, b): a \in A, b \in B\}$). For each pair, a similarity function is used that produces values between 0 and 1, where 1 represents maximum similarity and means that two data points are identical. Thus, for each de-identified record $b \in B$, the distance-based record linkage finds an original record $a \in A$ that minimises the distance to $b$, given the similarity function. 


To limit the risk of disclosure, appropriate PPTs are applied. There are two main categories of PPTs: non-perturbative and perturbative. We use the best-known examples of non-perturbative techniques: global re-coding (or generalisation) and suppression, which aim to reduce the amount of information in the original data. Global re-coding groups values into broader categories, while suppression replaces values with a missing value (NA) or a special character (*, ?). For the perturbative category, we use noise, which, unlike non-perturbative techniques, distorts the original data by adding or subtracting random values from the original value. 


To proceed with the de-identification process, it is essential to acknowledge that the team representing the data controller (healthcare professionals) is responsible for addressing the \textit{Five Safes} framework and conducting a comprehensive risk-benefit assessment.
This preliminary analysis is important to assess the relevance of the study and the associated risks from data collection to publication, before starting any data transformation procedures. In summary, the benefits of data sharing for this particular study include:
\begin{itemize}
\item simplification of the process of data availability, allowing rapid, secure access by healthcare researchers, and
\item liberate the information systems teams and the Data Protection Officer from recurrent requests for access to the same data set.
\end{itemize}

On the other hand, given the sensitive nature of patient data, it is imperative to mitigate data protection risks while ensuring the fundamental right to privacy of patients. Therefore, the establishment of a pseudonymisation methodology within data structures intended for research purposes becomes indispensable, allowing the risk of patient re-identification to be objectively quantified. Accordingly, the decision to release the data to the general public or to specific recipients is of great importance.

\subsection*{Data}
The provided data concerns Covid-19 cases in Portugal, namely, hospitalised cases from March 2020 to January 2021. The data is comprised of 1.716 individuals and 38 attributes. Such attributes include, for instance, the individuals' age (with an average of 67 years), date of first positive Covid-19 test, date of hospitalisation and discharge, the outcome of the case (indicating recovery or death) and several indicators of other pathologies. The definition of the attributes of this data set was based on the same attributes of the data set mentioned above, which was made available to researchers by the DGS in 2020 and proved to be not only not of limited quality~\cite{costa2021covid}, but also with the possibility of patient re-identification without much effort~\cite{carvalho2021fundamental}.
To better understand the data, we present some of the attribute values and their description in Table~\ref{tab:data}.

\begin{table}[ht!]
\centering
\scriptsize
\begin{adjustbox}{width=.8\textwidth}
\begin{tabular}{@{}lll@{}}
\toprule
\textbf{Attribute}                    & \textbf{Value}          & \textbf{Description}                        \\ \midrule
\textit{RecordId}                     & numeric                     & Record ID number                            \\
\textit{Age}                          & numeric                      & Patient's age                               \\
\textit{AgeDay}                       & numeric                   & Patient's age in days                       \\
\textit{AgeMonth}                     & numeric                     & Patient's age in months                     \\
\textit{CloseContactRecordId}         & numeric ID code                     & Record ID that was close to the positive case \\
\textit{DateOfFirstPositiveLabResult} & date YYYY/MM/DD HH:MM:SS & Date of first positive test                 \\
\textit{DateOfHospitalisation}        & date YYYY/MM/DD              & Date of hospitalisation                     \\
\textit{DateOfDischarge}              & date YYYY/MM/DD              & Date of discharge                           \\
\textit{DateOfOnset}                  & date YYYY/MM/DD                     & Date of onset                               \\
\textit{Gender}                       & F / M / Unknown                       & Patient's gender                            \\
\textit{Hospitalisation}              & Y / N / Unknown                       & The patient was hospitalised or not         \\
\textit{IntensiveCare}                & Y / N / Unknown                       & The patient was in intensive care or not    \\
\textit{Outcome}                      & H / D / O / N                        & Result of the case                          \\
\textit{PlaceOfInfection}             & Place code / Unknow                     & Place of infection                          \\
\textit{CANC}                         & Y / N /Unknown                       & The patient has cancer or not               \\ \bottomrule
\end{tabular}
\end{adjustbox}
\caption{Description of provided data.}
\label{tab:data}
\end{table}

We add value to the data set by analysing the distribution of values and performing feature engineering. 
Crucially, the validation of these new attributes is undertaken in collaboration with epidemiologists and bio-statisticians to ensure their clinical relevance. We create the attribute \textit{HospitalisationDays} which indicates the number of days that an individual was hospitalised by subtracting to the \textit{DateOfDischarge} the \textit{DateOfHospitalisation}. Then, we delete these two attributes.
Additionally, some attributes are excluded from the dataset. Although the \textit{Hospitalisation} attribute is very relevant in the context of the pandemic, this attribute takes the same value, since this this data set is only about the hospitalised patients. As such, this attribute is of no statistical interest. Others, such as \textit{CloseContactRecordId}, \textit{DateOfOnset} and \textit{PlaceOfInfection} have no statistical interest as there were many cases with missing information.

\section*{Results}
\subsection*{Initial Disclosure Risk}
In this study, we perform several assumptions regarding QIs.
We consider as QIs the attributes that have a high potential to increase the re-identification risk if combined with other attributes, for instance, \textit{Age}, \textit{DateOfFirstPositiveLabResult}, \textit{HospitalisationDays}, \textit{Gender} and \textit{Outcome}. Table~\ref{tab:risk} shows the percentage of re-identification risk for several scenarios, i.e. combination of QIs, determined by the percentage of the number of equivalence classes of size 1 ($k=1$) in the data set. An apparent result is the increased risk when we increase the set of selected QIs. We also observe that the date attribute has a strong influence on the risk. This is due to the granularity of the data, i.e. dates in the format "YYYY/MM/DD HH:MM:SS" are too aggressive for privacy, as they provide too much detail, which increases the number of unique cases. 

\begin{table}[ht!]
\centering
\scriptsize
\begin{adjustbox}{width=0.8\textwidth}

\begin{tabular}{ccccc|c}
\toprule
\multicolumn{5}{c|}{\textbf{QI attributes}}                  & \multirow{2}{*}[-3pt]{\textbf{\begin{tabular}[c]{@{}c@{}}Re-identification\\ risk (\%)\end{tabular}}} \\ \cline{1-5}

\textit{Age} & \textit{DateOfFirstPositiveLabResult} & \textit{HospitalisationDays} & \textit{Gender} & \multicolumn{1}{c|}{\textit{Outcome}} &                                              \\ \midrule
\checkmark   &                     &                      & \checkmark      &         & 1.86                                         \\
\checkmark   &                     &                      & \checkmark      & \checkmark       & 7.98                                         \\
\checkmark   &                     & \checkmark                    & \checkmark      &         & 64.74                                        \\
\checkmark   &                     & \checkmark                    & \checkmark      & \checkmark       & 79.14                                        \\
\checkmark   & \checkmark                   &                      & \checkmark      &         & 96.33                                        \\
\checkmark   & \checkmark                   &                      & \checkmark      & \checkmark       & 97.38                                        \\
\checkmark   & \checkmark                   & \checkmark                    & \checkmark      & \checkmark       & 99.77                                        \\ \bottomrule
\end{tabular}
\end{adjustbox}
\caption{Initial re-identification risk based on the selected quasi-identifiers.}
\label{tab:risk}
\end{table}

An important preliminary analysis in the evaluation of privacy risks is the selection of certain subsets where an attacker is assumed to have a high degree of certainty about specific background knowledge. In such a case, we assume that an attacker knows, for example, that a person has died, has been in a nursing home, has been in intensive care, or whether the case is related to newborns. Table~\ref{tab:subset_risk} shows the risk analysis for such subsets. We observe an increase in the risk when the subsets have a smaller number of observations. This is a normal result as the small subsets have a lower number of individuals sharing the same information and therefore a high number of unique cases. For the newborn subset, we use age in months instead of age in years. To prevent re-identification in this subset, we should remove the attributes \textit{AgeDay} and \textit{AgeMonth}. As the risk is also very critical for the nursing home subset, further protection should be applied.

\begin{table}[ht!]
\centering
\scriptsize
\begin{adjustbox}{width=0.8\textwidth}
\begin{tabular}{c|c|ccccc|c}
\toprule
\multirow{2}{*}[-3pt]{\textbf{Subset}} & \multirow{2}{*}[-3pt]{\textbf{Nr observations}} & \multicolumn{5}{c|}{\textbf{QI attributes}}                                  & \multirow{2}{*}[-3pt]{\textbf{\begin{tabular}[c]{@{}c@{}}Re-identification \\ risk (\%)\end{tabular}}} \\ \cmidrule(lr){3-7}
                                 &                                           & \textit{Age} & \textit{AgeMonths} & \textit{DateOfFirstPositiveLabResult} & \textit{HospitalisationDays} & \textit{Gender} &                                                       \\ \midrule
\multirow{3}{*}{Death}           & \multirow{3}{*}{368}                      & \checkmark   &               &                     &                      & \checkmark      & 7.61                                                  \\
                                 &                                           & \checkmark   &               &                     & \checkmark                    & \checkmark      & 85.6                                                  \\
                                 &                                           & \checkmark   &               & \checkmark                   &                      & \checkmark      & 100                                                   \\ \midrule
\multirow{3}{*}{Nursing home}    & \multirow{3}{*}{41}                       & \checkmark   &               &                     &                      & \checkmark      & 70.73                                                 \\
                                 &                                           & \checkmark   &               &                     & \checkmark                    & \checkmark      & 100                                                   \\
                                 &                                           & \checkmark   &               & \checkmark                   &                      & \checkmark      & 100                                                   \\ \midrule
\multirow{3}{*}{Intensive care}  & \multirow{3}{*}{529}                      & \checkmark   &               &                     &                      & \checkmark      & 5.67                                                  \\
                                 &                                           & \checkmark   &               &                     & \checkmark                    & \checkmark      & 89.41                                                 \\
                                 &                                           & \checkmark   &               & \checkmark                   &                      & \checkmark      & 100                                                   \\ \midrule
\multirow{3}{*}{Newborn}         & \multirow{3}{*}{12}                       &     & \checkmark             &                     &                      & \checkmark      & 41.67                                                 \\
                                 &                                           &     & \checkmark             &                     & \checkmark                    & \checkmark      & 100                                                   \\
                                 &                                           &     & \checkmark             & \checkmark                   &                      & \checkmark      & 100                                                   \\ \bottomrule
\end{tabular}
\end{adjustbox}
\caption{Re-identification risk in subsets based on the selected quasi-identifiers.}
\label{tab:subset_risk}
\end{table}

As this data set provides information on several pathologies, special attention should be paid to these subsets, as certain pathologies are only common to a small number of people, e.g. people with HIV, pregnant women or even obese young men. These cases can easily be re-identified by an attacker simply by accessing social networks where people share pictures. Henceforth, a strong recommendation for similar cases is to group such attributes into one that we can call, "Other pathologies" for example.

In the de-identification process, utility is usually measured using the raw data. However, this applies when the purpose of the release is for example for modelling tasks. In our case, the utility was measured at each iteration of the PPTs, evaluating the impact of the transformations on the data set.

\subsection*{Effectiveness of Privacy-Preserving Techniques}

During data analysis, we noticed some re-incidents, referring to people who were hospitalised more than once. However, such cases represent a very small part of the data and may be potentially relevant to an attacker. Therefore, we apply suppression at the row level by retaining only the first date of hospitalisation for re-incident cases due to early discharge. This transformation results in a lower number of observations and, consequently, a higher re-identification risk. The dataset contains 1.685 individuals, and Table~\ref{tab:reassess_risk} provides the reassessment of the disclosure risk. All scenarios have increased the disclosure risk compared to Table~\ref{tab:risk}. Henceforth, we compare the re-identification risk of the posterior data transformations with these results.

\begin{table}[ht!]
\centering
\scriptsize
\begin{adjustbox}{width=0.8\textwidth}
\begin{tabular}{ccccc|c}
\toprule
\multicolumn{5}{c|}{\textbf{QI attributes}}                  & \multirow{2}{*}[-3pt]{\textbf{\begin{tabular}[c]{@{}c@{}}Re-identification\\ risk (\%)\end{tabular}}} \\ \cline{1-5}

\textit{Age} & \textit{DateOfFirstPositiveLabResult} & \textit{HospitalisationDays} & \textit{Gender} & {\textit{Outcome}}  &                                   \\ \midrule
\checkmark   &                     &                      & \checkmark      &         & 1.9                                         \\
\checkmark   &                     &                      & \checkmark      & \checkmark       & 8.07                                         \\
\checkmark   &                     & \checkmark                    & \checkmark      &         & 65.28                                        \\
\checkmark   &                     & \checkmark                    & \checkmark      & \checkmark       & 79.76                                        \\
\checkmark   & \checkmark                   &                      & \checkmark      &         & 99.88                                        \\
\checkmark   & \checkmark                   &                      & \checkmark      & \checkmark       & 100                                        \\
\checkmark   & \checkmark                   & \checkmark                    & \checkmark      & \checkmark       & 100                                        \\ \bottomrule
\end{tabular}
\end{adjustbox}
\caption{Re-assessment of the re-identification risk based on the selected quasi-identifiers.}
\label{tab:reassess_risk}
\end{table}

To reduce the re-identification risk, we focus on attributes with dates and hours as these are very impacting in the re-identification risk. We start by removing the hours to reduce the level of detail. In each iteration, it is important to re-assess the risk of re-identification to verify the impact that the transformation has had on privacy risk. As a result, the risk of re-identification drops to 89.02\% when we consider age, date of test result and gender. However, there is still a high percentage of singling out cases that should be addressed.

\begin{figure}[ht!]
  \centering
    \includegraphics[width=0.6\textwidth]{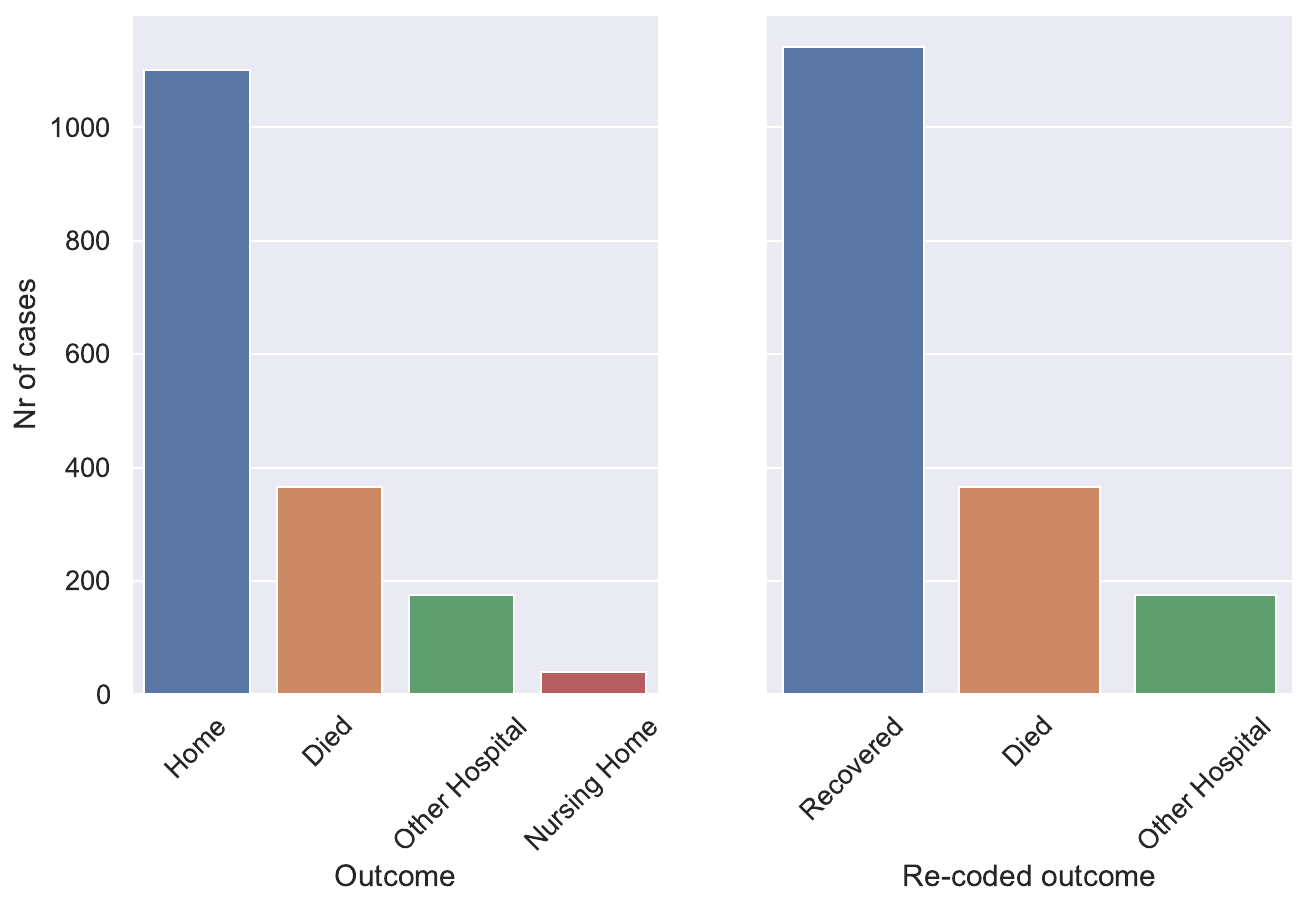}
    \caption{Discharge destiny of the initial Covid-19 cases (left) and after the re-coding (right).}
    \label{fig:outcome}
\end{figure}

As we have previously observed in the subsets analysis, there are a few individuals who have gone to a nursing home, and as such a case corresponds to old people, it requires special attention. Furthermore, we highlight the importance of minimal information sharing for privacy protection, which can be solved by aggregating nursing home to home cases. We call such an aggregation "Recovered". Figure~\ref{fig:outcome} shows the distribution of individuals concerning outcome cases for the original data and after this transformation. The risk remains the same for the QIs combination that includes \textit{HospitalisationDays} or \textit{DateOfFirstPositiveLabResult}, as this transformation has no effect when combined with such attributes. On the other hand, considering \textit{Age}, \textit{Gender} and the updated \textit{Outcome}, the risk drops from 8\% to 6.5\%.

\begin{figure}[ht!]
  \centering
    \includegraphics[width=0.6\textwidth]{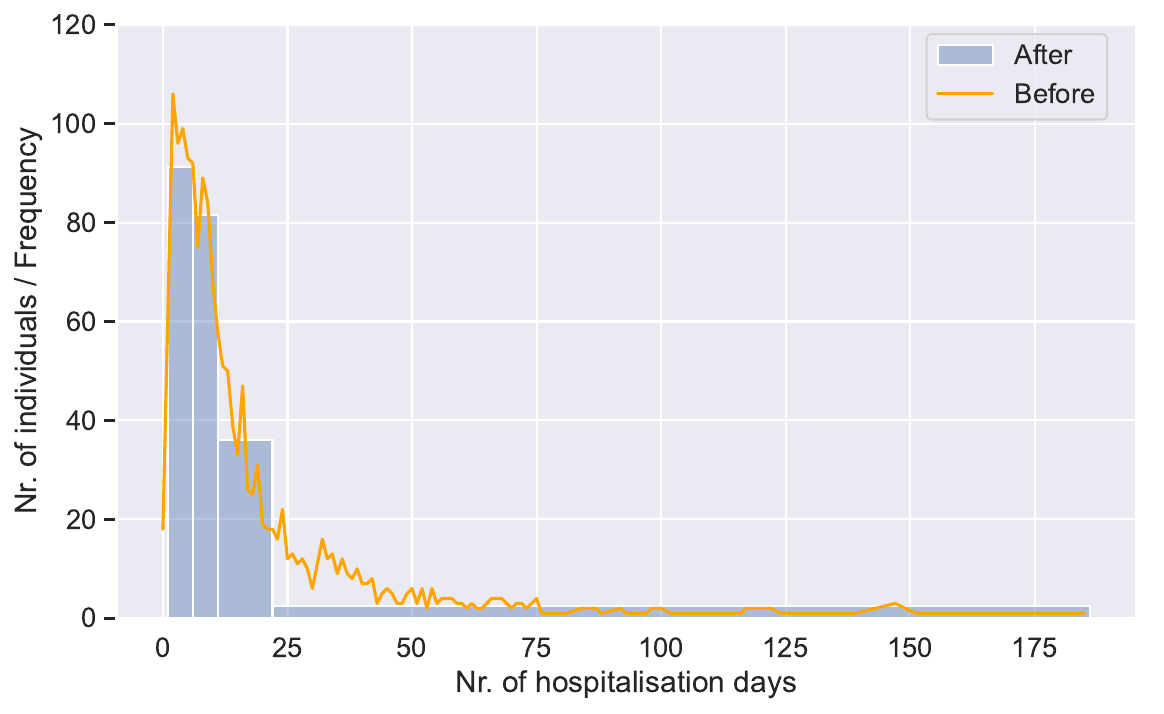}
    \caption{Distribution of hospitalisation days concerning the number of individuals compared to the frequency of individuals (number of observations divided by the bin width) after the generalisation to quartiles.}
    \label{fig:hospdays}
\end{figure}

From Table~\ref{tab:reassess_risk}, we can verify that the re-identification risk increases substantially with the addition of the attribute on hospitalisation days. At this point, it was really interesting to see that once the epidemiologist and bio-statistician understood the PPT considered before, they contributed actively by proposing to represent this attribute in quartiles, as the level of statistical utility remains appropriate to the goal. This approach drastically reduced the risk of re-identification from 65.28\% to 8.7\%. This was an important takeout of this process: it is of utmost importance to raise the PPTs awareness of the end user, as once they understand it they take the lead and propose new methods that do not compromise the utility of the data.
Such a transformation is illustrated in Figure~\ref{fig:hospdays}, where we show how many days certain individuals were hospitalised. For this purpose, we present the distribution of the original data and the frequency of the observations after applying the quartiles. We can observe that this generalisation allows us to maintain the distribution close to the original, and therefore preserve the utility.

\begin{figure}[ht!]
  \centering
    \includegraphics[width=\textwidth]{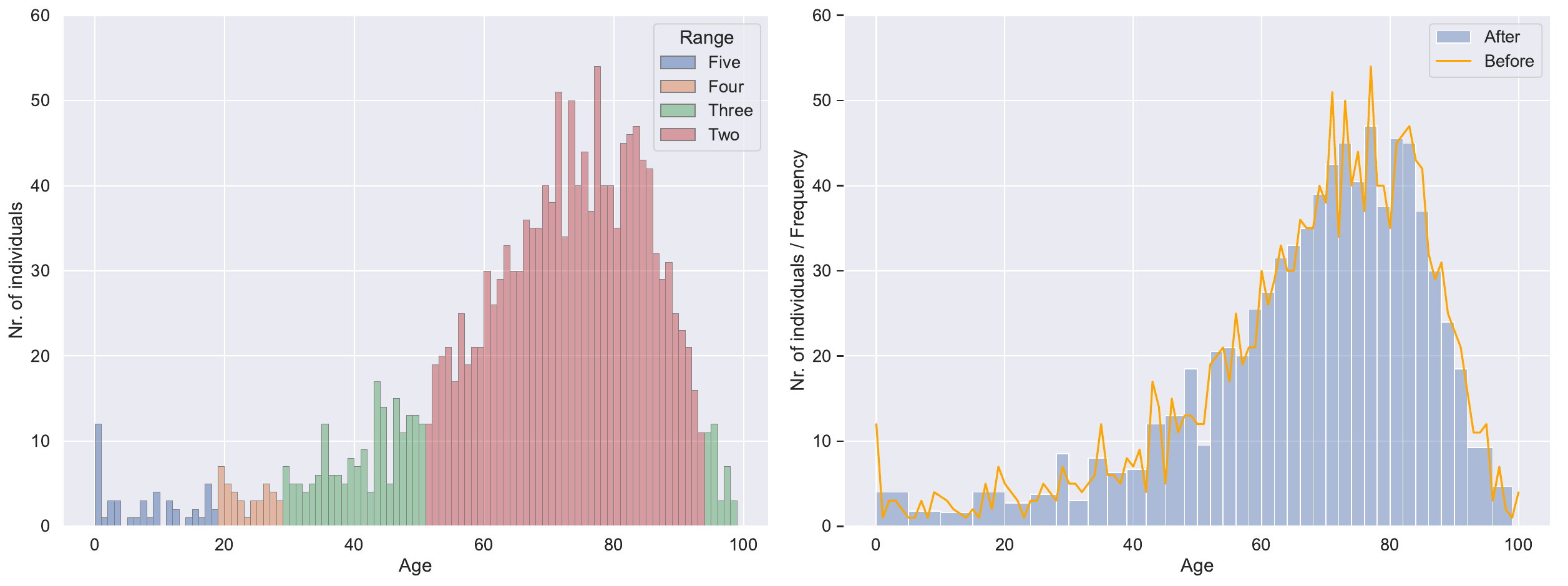}
    \caption{Distribution of \textit{Age} attribute with the respective representation of dynamic ranges (left) and the comparison of original distribution with the transformed data using such ranges (right) in terms of frequencies.}
    \label{fig:age_ranges}
\end{figure}

To further reduce the risk of re-identification, we can also generalise the age. An approach that preserves more utility is to generalise across different intervals. In general, fewer intervals are applied at the top of the curve (more individuals in the same age range) and fewer intervals at the tails of the curve. Figure~\ref{fig:age_ranges} shows an example applied with different ranges (left). In order to make the newborns indistinguishable from other individuals, we have aggregated them to the following ages. The figure on the right shows the effect of the transformation using the dynamic ranges.

\begin{figure}[ht!]
  \centering
    \includegraphics[width=0.6
    \textwidth]{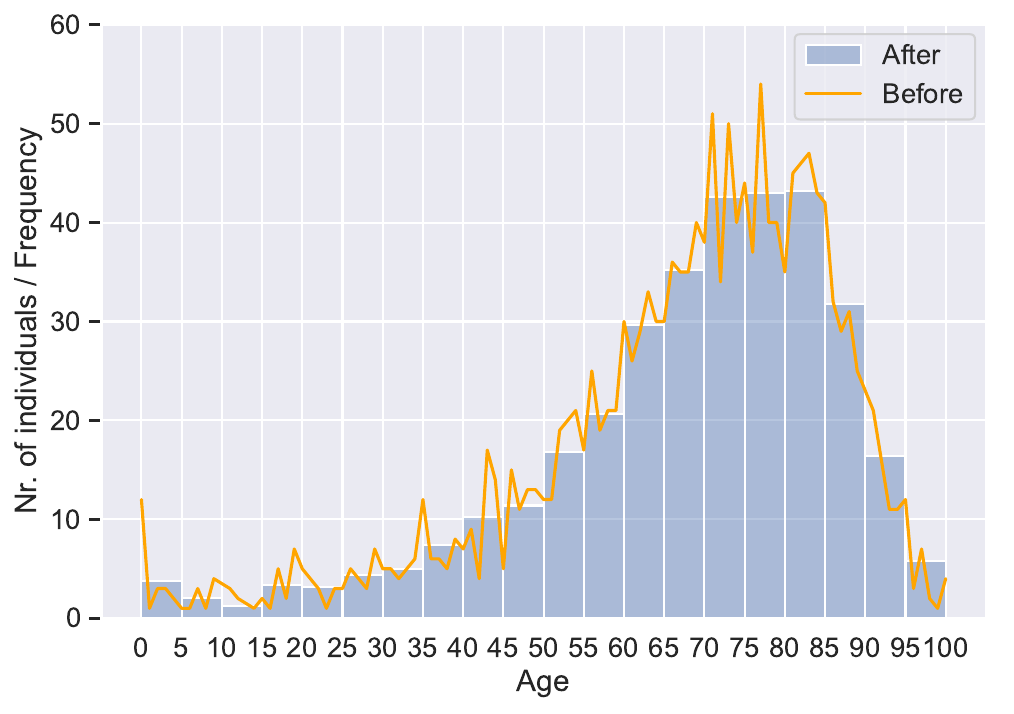}
    \caption{Distribution of ages concerning the number of individuals compared to the frequency of individuals after the generalisation to intervals of 5.}
    \label{fig:age5}
\end{figure}

In addition to maintaining a high utility, we obtain a re-identification risk of 0\% for the combination of \textit{Age} and \textit{Gender}. Similarly, adding the recoded result reduces the risk from 6.5\% to 1.8\%. However, this approach has not been further developed, because to add more knowledge to the data set, it is essential to combine it with other data sets. The bio-statistician argued that in general, open data platforms for open science in Portugal present the age in intervals of 5 years. For this reason, in Figure~\ref{fig:age5} we provide the respective frequency of ages considering such an interval. Although the level of risk is maintained, the granularity of the data is reduced.

At this point, only further processing of the date type attribute remains to achieve a reasonable level of re-identification risk. To this end, we start the experiments by applying generalisation over periods of days in the date of the first positive case. Table~\ref{tab:dates_gen} shows three examples of such periods for \textit{DateOfFirstPositiveLabResult} attribute regarding the QIs. It should be noted that the results are sequential, i.e. we present the results with the previous transformations. Thus, these results refer to the generalisation of the date of the first positive case, re-coded outcome, days of hospitalisation in quartiles and generalisation of age in intervals of 5 years.

\begin{table}[ht!]
\centering
\scriptsize
\begin{adjustbox}{width=0.8\textwidth}

\begin{tabular}{@{}ccccc|c|c@{}}
\toprule
\multicolumn{5}{c|}{\textbf{QI atrributes}}                                                                                                             & \multirow{2}{*}[-3pt]{\textbf{Generalisation}} & \multirow{2}{*}[-3pt]{\textbf{\begin{tabular}[c]{@{}c@{}}Re-identification\\ risk (\%)\end{tabular}}} \\ \cmidrule(r){1-5}
\textit{Age}              & \textit{DateOfFirstPositiveLabResult} & \textit{HospitlisationDays} & \textit{Gender}           & \textit{Outcome}          &                                          &                                                                                                 \\ \midrule
\checkmark & \checkmark             &    &        \checkmark                   &                           & \multirow{3}{*}{2 days}                  & 42.61                                                                                           \\
\checkmark & \checkmark             &                             & \checkmark & \checkmark &                                          & 62.01                                                                                           \\
\checkmark & \checkmark             & \checkmark   & \checkmark & \checkmark &                                          & 84.28                                                                                           \\ \midrule
\checkmark & \checkmark             &    &      \checkmark                     &                           & \multirow{3}{*}{5 days}                  & 23.20                                                                                           \\
\checkmark & \checkmark             &                             & \checkmark & \checkmark &                                          & 38.57                                                                                           \\
\checkmark & \checkmark             & \checkmark   & \checkmark & \checkmark &                                          & 69.44                                                                                           \\ \midrule
\checkmark & \checkmark             &    &      \checkmark                     &                           & \multirow{3}{*}{7 days}                  & 19.58                                                                                           \\
\checkmark & \checkmark             &                             & \checkmark & \checkmark &                                          & 32.58                                                                                           \\
\checkmark & \checkmark             & \checkmark   & \checkmark & \checkmark &                                          & 63.68                                                                                           \\ \bottomrule
\end{tabular}
\end{adjustbox}
\caption{Re-identification risk results concerning diverse generalisation transformation for the date of first positive test.}
\label{tab:dates_gen}
\end{table}

As mentioned above, by removing the hour from this attribute we obtained an 89.02\% risk of re-identification for \textit{Age}, \textit{DateOfFirstPositiveLabResult} and \textit{Gender}. Although using a 2 day interval results in a considerable reduction, 42.6\% is still a cause for concern. Even considering a 7-day interval is not a satisfactory result. Thus, we need intervals longer than 7 days, which may undermine the utility. 
Given this study is over a short period, and the dates of the test results are important for further understanding of the Covid-19 cases, we need to preserve as much information as possible. A plausible solution that is best suited for this purpose is the addition of noise. Such an approach will distort the dates, making them untruthful, but this ensures that there is no loss of granularity, as several records have the date very close to the original. We therefore apply noise between the interval [-3, 3], which means that for each date a random value is added from this range. 

Once the noise is introduced, $k$ anonymity is inappropriate as this transformation does not involve changing the granularity of the data. In this case, the equivalence classes including the \textit{DateOfFirstPositiveLabResult} attribute will be equivalent with or without noise. Therefore, we use record linkage to address such a shortcoming, which allows us to determine how many records in the transformed data set are identical to the original. This comparison is made using only QIs to limit the comparison space. Table~\ref{tab:record_linkage} shows the results after introducing noise in the values of the date attribute. 

\begin{table}[ht!]
\centering
\scriptsize
\begin{adjustbox}{width=0.8\textwidth}
\begin{tabular}{@{}ccccc|c@{}}
\toprule
\multicolumn{5}{c|}{\textbf{QI atrributes}}                                                                                                             & \multirow{2}{*}[-3pt]{\textbf{\begin{tabular}[c]{@{}c@{}}Re-identification\\ risk (\%)\end{tabular}}} \\ \cmidrule(r){1-5}
\textit{Age}              & \textit{DateOfFirstPositiveLabResult} & \textit{HospitlisationDays} & \textit{Gender}           & \textit{Outcome}          &                                                                                                 \\ \midrule
\checkmark & \checkmark             &                             & \checkmark &                           & 30                                                                                           \\
\checkmark & \checkmark             &                             & \checkmark & \checkmark & 0                                                                                               \\
\checkmark & \checkmark             & \checkmark   & \checkmark & \checkmark & 0                                                                                               \\ \bottomrule
\end{tabular}
\end{adjustbox}
\caption{Re-identification risk results using record linkage after the addition of noise in the date of first positive test.}
\label{tab:record_linkage}
\end{table}

\begin{figure}[!ht]
  \centering
    \includegraphics[width=0.7
    \textwidth]{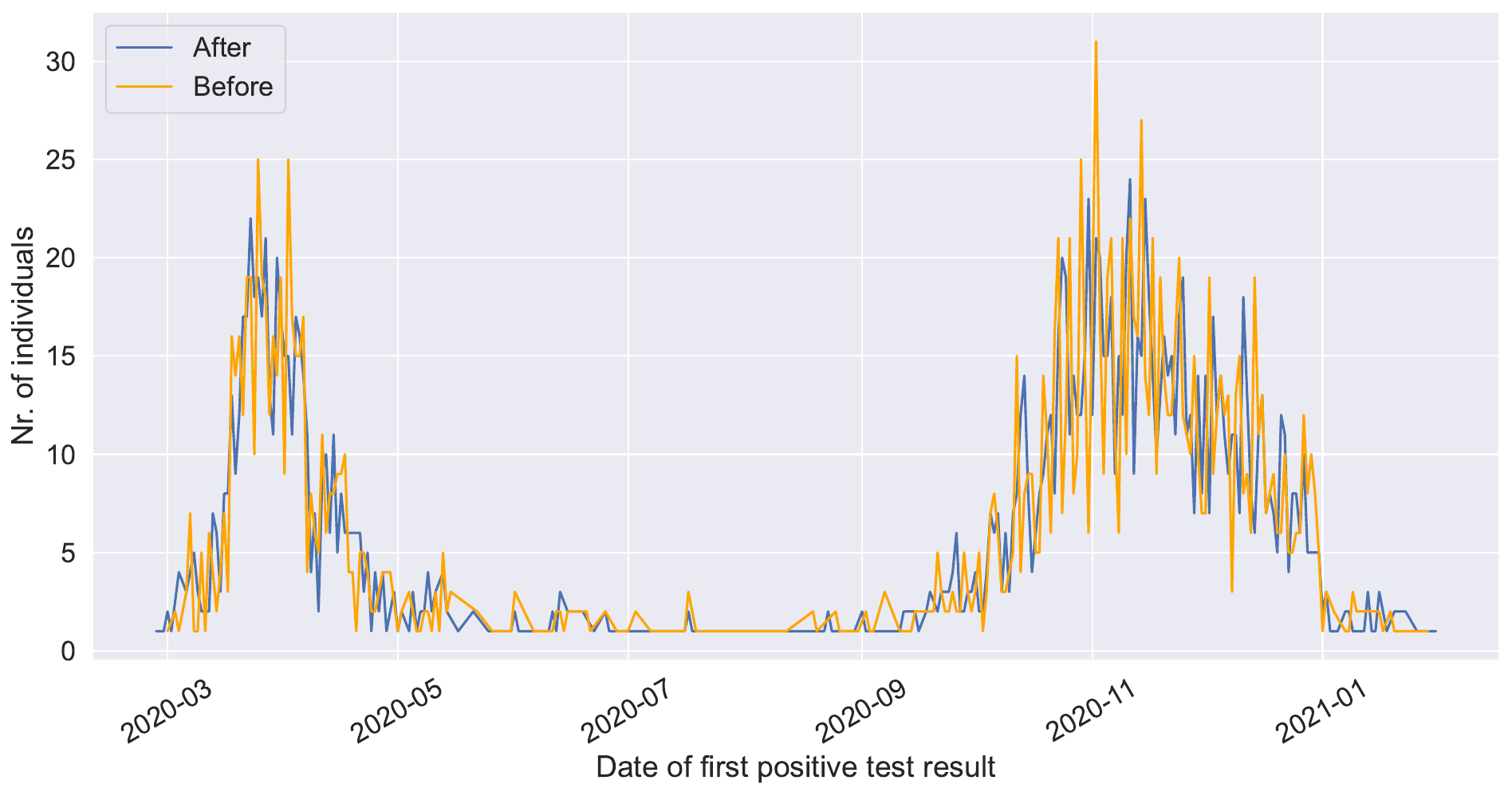}
    \caption{Number of individuals that obtained positive test results along with the distribution of positive tests after the addition of noise within the range [-3, 3].}
    \label{fig:noise}
\end{figure}

Contrary to the results with $k$-anonymity, the more added noise, the lower the risk of re-identification, as an attacker is less likely to have the exact information of all QIs. However, record linkage does not evaluate whether a matched pair is a true match or not. A false match is also critical from a privacy perspective, as misleading information about an individual may be released. We noticed that 17\% are false positives, which means that despite a match rate of about 30\%, the margin of error is over 50\%.
Regarding the utility with the noise addition, Figure~\ref{fig:noise} shows the distribution before and after the transformation on \textit{DateOfFirstPositiveLabResult} attribute, and it is visible that the two distributions remain very similar.

The final results are presented in Table~\ref{tab:final} with the corresponding PPTs and their parameterisation. We present the results with the two metrics, record linkage when QIs contain attributes with noise, and $k$-anonymity otherwise. With the iterations between the privacy and the health experts, it was possible to reduce the re-identification risk to 0\% for some scenarios with a high utility, and therefore no further transformations will be performed on the data. 

\begin{table}[!ht]
\centering
\scriptsize
\begin{adjustbox}{width=\textwidth}
\begin{tabular}{@{}cccccc|c|c@{}}
\cmidrule(l){2-8}
{\color[HTML]{656565} \textbf{Transformation}} & \begin{tabular}[c]{@{}c@{}}Generalisation\\ (range 5)\end{tabular} & \begin{tabular}[c]{@{}c@{}}Noise\\ ({[}-3, 3{]})\end{tabular} & \begin{tabular}[c]{@{}c@{}}Generalisation\\ (centrality - quartiles)\end{tabular} & -                         & \begin{tabular}[c]{@{}c@{}}Generalisation\\ (re-coding category)\end{tabular} &                                                                                                  &                                   \\ \cmidrule(lr){2-6}
{\color[HTML]{656565} \textbf{QI attributes}}  & \textit{Age}                                                       & \textit{DateOfFirstPositiveLabResult}                       & \textit{HospitalisationDays}                                                      & \textit{Gender}           & \textit{Outcome}                                                              & \multirow{-3}{*}{\textbf{\begin{tabular}[c]{@{}c@{}}Re-identification\\ risk (\%)\end{tabular}}} & \multirow{-3}{*}{\textbf{Metric}} \\ \cmidrule(l){2-8} 
                                               & \checkmark                                          &                                                             &                                                                                   & \checkmark &                                                                               & 0                                                                                                & $k$-anonymity                     \\
                                               & \checkmark                                          &                                                             &                                                                                   & \checkmark & \checkmark                                                     & 0.89                                                                                             & $k$-anonymity                     \\
                                               & \checkmark                                          &                                                             & \checkmark                                                         & \checkmark &                                                                               & 1.07                                                                                             & $k$-anonymity                     \\
                                               & \checkmark                                          &                                                             & \checkmark                                                         & \checkmark & \checkmark                                                     & 3.38                                                                                             & $k$-anonymity                     \\
                                               & \checkmark                                          & \checkmark                                   &                                                                                   & \checkmark &                                                                               & 13                                                                                               & Record linkage                    \\
                                               & \checkmark                                          & \checkmark                                   &                                                                                   & \checkmark & \checkmark                                                     & 0                                                                                                & Record linkage                    \\
                                               & \checkmark                                          & \checkmark                                   & \checkmark                                                         & \checkmark & \checkmark                                                     & 0                                                                                                & Record linkage                    \\ \cmidrule(l){2-8} 
\end{tabular}
\end{adjustbox}
\caption{Final re-identification risk results after the application of privacy-preserving techniques to the quasi-identifiers attributes.}
\label{tab:final}
\end{table}

Nevertheless, as initially mentioned, it is important to analyse subsets, as an attacker's knowledge may be underestimated. Therefore, Table~\ref{tab:pathologies} shows the impact of the final transformations on the re-identification risk when the attacker knows in advance the age, gender and pathology associated with a specific individual. Of the previous QIs attributes, we only consider \textit{Age} and \textit{Gender} for this analysis, as these attributes are generally the easiest for an attacker to obtain. We observe that all subsets maintain the re-identification risk close to 0\%.

\begin{table}[!ht]
\centering
\scriptsize
\begin{adjustbox}{width=0.4\textwidth}
\begin{tabular}{{l|cc}}
\toprule
\multirow{2}{*}[-3pt]{\textbf{Pathology}} & \multicolumn{2}{c}{\textbf{Re-identification risk (\%)}} \\ \cmidrule(l){2-3} 
                                    & \multicolumn{1}{c|}{Original}         & Protected        \\ \midrule
Cancer                              & \multicolumn{1}{c|}{4.27}             & 0.47             \\
\textit{Cerebrovascular}            & \multicolumn{1}{c|}{3.56}             & 0.18             \\
\textit{Diabetes}                   & \multicolumn{1}{c|}{3.44}             & 0.12             \\
\textit{Kidney}                     & \multicolumn{1}{c|}{3.5}              & 0.18             \\
\textit{Liver}                      & \multicolumn{1}{c|}{4.09}             & 0.18             \\
\textit{Lung}                       & \multicolumn{1}{c|}{3.15}             & 0.42             \\
\textit{Heart}                      & \multicolumn{1}{c|}{3.2}              & 0.3              \\
\textit{Smoking}                    & \multicolumn{1}{c|}{3.98}             & 0.47             \\
\textit{Obesity}                    & \multicolumn{1}{c|}{4.21}             & 0.12             \\ \bottomrule
\end{tabular}
\end{adjustbox}
\caption{Impact of re-identification risk in subsets concerning \textit{Age}, \textit{Gender} and a certain pathology after the application of privacy-preserving techniques.}
\label{tab:pathologies}
\end{table}

\section*{Discussion}
Open data has long been used to benefit society in a variety of domains. The pandemic has highlighted the urgency of making data available to the public. To achieve this, the privacy of individuals should be a priority. However, to ensure privacy, data are often minimised and generalised to the point of being statistically useless. This is a consequence of misuse of optimal PPTs or high PPT parameters, use of misleading case studies, disregard of the purpose of releasing the data, and many other causes.

Nowadays, privacy faces three main challenges: unawareness of the importance of protecting private information, lack of knowledge about privacy preservation methods, and the idea that preserving privacy destroys utility, which has long been nurtured but needs to be deconstructed. It is crucial to convey that although the process of de-identification requires extra effort, it is possible to preserve privacy without compromising utility too much. 

One of the main tasks for successful de-identification is the cleaning of the data set to be protected. Some criticisms about the quality of the data provided for open data on Covid-19 cases in Portugal have already been pointed out~\cite{costa2021covid}. If we use a data set of very poor quality, privacy transformations will also produce data of equal or worse quality. It is very important to perform a pre-analysis, remove statistically insignificant attributes and aggregate cases to minimise information. Once the healthcare professionals understood the purpose of the de-identification, they participated in all steps, including providing several suggestions to improve privacy while maintaining utility. This iteration between the privacy and healthcare teams resulted in a faster and more efficient process. 

Despite the successful outcome, there are additional concerns that need to be addressed. We focus on single out cases as they pose the greatest risk. However, one person sharing information with another means that each has a 50\% chance of being re-identified. In our study, the PPTs applied ensured that no individual was singled out or shared information with another. Nevertheless, 3 individuals share the same information, posing a re-identification risk of 33\%. Since we guarantee a minimum of equivalence classes of size 3, we consider such an outcome reasonable for our particular case. To achieve higher equivalence classes, the PPT parameters must be readjusted or other techniques must be used in combination.



Furthermore, we focus only on identity disclosure. However, it should be considered whether there are cases of possible attribute disclosure. 
For instance, we found that several women of similar age have cerebrovascular problems.
Although it is not possible to directly identify each of these women, we can conclude that they all suffer from the same pathology. This means that there is a loss of privacy, namely the disclosure of attributes. Typically, the regulator's main concern is to ensure that there is no entity disclosure. However, these concerns should not be ignored. Other metrics should be used to evaluate attribute disclosure, such as $l$-diversity~\cite{machanavajjhala2007diversity}. Also, we stress that the record linkage metric is useful for both re-identification and attribute disclosure~\cite{carvalho2021fundamental}. 
Although the two risk metrics used are applied for different purposes, namely $k$-anonymity on truthful and record linkage untruthful transformations, both metrics should be used simultaneously, since the former gives us the proportion of single outs, whereas the latter provides the proportion of information that is unprotected.

The final task after de-identification is to submit a report demonstrating the benefits of data release, including all de-identification steps and corresponding risk levels, to the supervisory authority, which will assess whether the risk is acceptable to proceed to open data release or data sharing (GDPR, Article 51). Nevertheless, all elements of the processing of personal data should be publicly described in a comprehensive overview. 

To conclude, we emphasise that de-identification experts should educate and promote awareness on how to protect private information. Collaboration with data custodians - individuals well-versed in the data domain - is essential. During this process, the experts should demonstrate the different stages of the de-identification process, as this is the first step in raising awareness of the workflow involved. 
The synergy between the data protection team, the data release team and the end users is crucial for producing a statistically significant data set with a high level of privacy protection. 

\section*{Data availability}
The data can be provided upon request. The data are from patients who have had Covid-19 and the hospital only provided the data for research under an ethics committee approval. The ethics committee that approved our study was the Faculty of Medicine of the University of Porto. We had access to the data through the collaboration between the institutions. 

\section*{Code availability}
The code can be provided upon request. 

\bibliography{sample}



\section*{Author contributions statement}

All authors contributed equally to the manuscript. 

\section*{Competing interests} 
The authors declare no competing interests.

\end{document}